\documentclass[useAMS,usenatbib]{mn2e}
\usepackage{graphicx,amsmath,amssymb,subfigure,rotating,psfig}

\def\simgt{\mathrel{\lower0.6ex\hbox{$\buildrel {\textstyle >} \over {\scriptstyle \sim}$}}}
\def\simlt{\mathrel{\lower0.6ex\hbox{$\buildrel {\textstyle <} \over {\scriptstyle \sim}$}}}

\newcommand{\gtsim}{\mbox{{\raisebox{-0.4ex}{$\stackrel{>}{{\scriptstyle\sim}}$}}}}
\newcommand{\ltsim}{\mbox{{\raisebox{-0.4ex}{$\stackrel{<}{{\scriptstyle\sim}}$}}}}

\newcommand{\apj}{ApJ}
\newcommand{\apjl}{ApJL}
\newcommand{\mnras}{MNRAS}
\newcommand{\aj}{AJ}
\newcommand{\apjs}{ApJS}
\newcommand{\nat}{Nature}
\newcommand{\araa}{ARA\&A}
\newcommand{\aap}{A\&A}

\newcommand{\pasp}{PASP}

\begin{document}

\title[The far-infrared--radio correlation at $z<0.5$]{{\em Herschel}-ATLAS: The far-infrared--radio correlation at $z<0.5$\thanks{Herschel is an ESA space observatory with science instruments provided by European-led Principal Investigator consortia and with important participation from NASA}}

\author[M.~J.~Jarvis et al.]
{Matt~J.~Jarvis$^{1}$\thanks{Email: M.J.Jarvis@herts.ac.uk},  D.J.B.~Smith$^{2}$,  D.G.~Bonfield$^{1}$,  M.J.~Hardcastle$^{1}$, J.T.~Falder$^{1}$,
\newauthor
J.A.~Stevens$^{1}$
~R.J.~Ivison$^{3}$, 
~R.~Auld$^4$,
~M.~Baes$^{5}$,
~I.K.~Baldry$^6$, 
~S.P.~Bamford$^2$, 
\newauthor
N.~Bourne$^{2}$,
~S.~Buttiglione$^7$,  
~A. Cava$^{8}$,
~A.~Cooray$^{9}$, 
~A.~Dariush$^4$,
~G.~de Zotti$^{7}$,
\newauthor
J.S.~Dunlop$^{10}$, 
~L. Dunne$^{2}$,
~S. Dye$^{4}$,
~S. Eales$^4$, 
~J.~Fritz$^{5}$, 
~D.T.~Hill$^{11}$,   
\newauthor
R.~Hopwood$^{12}$, 
~D.H.~Hughes$^{13}$, 
~E.~Ibar$^{3}$, 
~D.H.~Jones$^{14}$, 
~L.~Kelvin$^{11}$, 
\newauthor
A.~Lawrence$^{10}$,
~L.~Leeuw$^{15}$, 
~J.~Loveday$^{16}$, 
~S.J. Maddox$^{2}$,
~M.J.~Micha{\l}owski$^{10}$,
\newauthor
M.~Negrello$^{12}$,
~P.~Norberg$^{10}$, 
~M.~Pohlen$^4$,
~M.~Prescott$^{6}$, 
~E.E.~Rigby$^2$,
~A.~Robotham$^{11}$, 
\newauthor
G.~Rodighiero$^7$, 
~D.~Scott$^{17}$,
~R.~Sharp$^{14}$, 
~P.~Temi$^{15}$, 
~M.A.~Thompson$^{1}$,
\newauthor
P.~van der Werf$^{18}$,
~E.~van~Kampen$^{19}$, 
~C.~Vlahakis$^{18}$,
~G.~White$^{12,20}$
\\
$^{1}$Centre for Astrophysics, Science \&\ Technology Research
Institute, University of Hertfordshire, Hatfield, Herts, AL10 9AB,
UK \\
$^{2}$Centre for Astronomy and Particle Theory, School of
 Physics \&\ Astronomy, The University of Nottingham, University Park, Nottingham, \\NG7 1HR, UK \\
$^{3}$UK Astronomy Technology Centre, Royal Observatory, Edinburgh,
EH9 3HJ, UK\\
$^4$School of Physics \&\ Astronomy, Cardiff University, Queen
Buildings, The Parade, Cardiff, CF24 3AA, UK \\
$^{5}$Sterrenkundig Observatorium, Universiteit Gent, Krijgslaan 281
S9, B-9000 Gent, Belgium\\
$^6$Astrophysics Research Institute, Liverpool John Moores
University, Twelve Quays House, Egerton Wharf, Birkenhead, CH41 1LD,
UK\\
 $^7$INAF--Osservatorio Astronomico di Padova, Vicolo Osservatorio 5, I-35122,
Padova, Italy\\
$^8$Instituto de Astrof\'isica de Canarias (IAC) and Departamento de
 Astrof\'isica de La Laguna (ULL), La Laguna, Tenerife, Spain\\
$^{9}$Department of Physics and Astronomy, University of California, Irvine, CA 92697, USA \\
$^{10}$SUPA, Institute for Astronomy, University of Edinburgh, Royal
Observatory, Blackford Hill, Edinbugh EH9 3HJ, UK \\
$^{11}$SUPA, School of Physics and Astronomy, University of St.
Andrews, North Haugh, St. Andrews, KY16 9SS, UK\\
$^{12}$Department of Physics and Astronomy, The Open University, Walton Hall, Milton Keynes, MK7 6AA, UK\\
$^{13}$Instituto Nacional de Astrof\'isica \'Optica y Electr\'onica (INAOE), Aptdo. Postal 51 y 72000 Puebla, Mexico\\
$^{14}$Anglo-Australian Observatory, PO Box 296, Epping, NSW 1710, Australia\\
$^{15}$Astrophysics Branch, NASA Ames Research Center, Mail Stop 2456,  Moffett Field, CA 94035, USA\\
$^{16}$Astronomy Centre, Department of Physics and Astronomy, School
of Maths and Physical Sciences, Pevensey II Building, \\University of
Sussex, Falmer, Brighton, BN1 9QH, UK\\
$^{17}$Department of Physics and Astronomy, University of British Columbia, 6224 Agricultural Road, Vancouver, BC, V6T1Z1, Canada \\
$^{18}$Leiden Observatory, Leiden University, P.O. Box 9513, NL - 2300 RA Leiden, The Netherlands\\
$^{19}$European Southern Observatory, Karl-Schwarzschild-Strasse 2, D-85748, Garching bei M\"unchen, Germany\\
$^{20}$The STFC Rutherford Appleton Laboratory, Didcot, Oxfordshire, UK\\
\\
\\
\\
\\
\\
\\
\\
\\
\\
\\
\\
\\
\\
\\
\\
\\
\\
\\
}

\maketitle

\begin{abstract}
We use data from the {\em Herschel}-ATLAS to investigate the evolution of the far-infrared--radio correlation over the redshift range $0 < z< 0.5$. Using the total far-infrared luminosity of all $>5\sigma$ sources in the {\em Herschel}-ATLAS Science Demonstration Field and cross-matching these data with radio data from the Faint Images of the Radio Sky at Twenty-Centimetres (FIRST) survey and the NRAO VLA Northern Sky Survey (NVSS), we obtain 104 radio counterparts to the {\em Herschel} sources. With these data we find no evidence for evolution in the far-infrared--radio correlation over the redshift range $0<z<0.5$, where the median value for the ratio between far-infrared and radio luminosity, $q_{\rm IR}$, over this range is $q_{\rm IR} = 2.40\pm 0.12$ (and a mean of $q_{\rm IR}=2.52 \pm 0.03$ accounting for the lower limits), consistent with both the local value determined from {\em IRAS} and values derived from surveys targeting the high-redshift Universe.  By comparing the radio fluxes of our sample measured from both FIRST and NVSS we show that previous results suggesting an increase in the value of $q_{\rm IR}$ from high to low redshift may be the result of resolving out extended emission of the low-redshift sources with relatively high-resolution interferometric data, although AGN contamination could still play a significant role.

We also find tentative evidence that the longer wavelength, cooler dust is heated by an evolved stellar population which does not trace the star-formation rate as closely as the shorter wavelength  $\ltsim 250~\mu$m emission or the radio emission, supporting suggestions based on detailed models of individual galaxies.
\end{abstract}


\section{Introduction}

The relation between far-infrared emission and the radio emission from galaxies at all redshifts is surprisingly tight \citep{deJong85, Helou85, Condon92, Garret02} and leads to the conclusion that both trace recent star-formation, in the local and distant Universe. The far-infrared emission is believed to arise  from the thermal emission of dusty clouds surrounding regions of star formation, whereas the radio emission arises from cosmic-ray electrons accelerated in supernova remnants of the dying stars, which emit synchrotron radiation. 

It is however unclear why there should be such a correlation between the thermal far-infrared emission and the non-thermal radio emission over such a wide range of galaxy types and masses, from starburst systems to more normal galaxies. As a result of this, many models resort to a relatively significant amount of fine tuning, such as assuming a much stronger magnetic field than what is estimated via minimum energy arguments \citep[e.g.][]{Thompson06}. A full discussion of such arguments is beyond the scope of this paper; however, we refer to the reader to 
\citet{Lacki1} and \citet{Lacki2} who provide a detailed discussion of the various physical interpretations of the far-infrared--radio correlation (FIRC).

Observationally, recent work has concentrated on exploring the FIRC as a function of redshift, mainly because of the preponderance of deep {\em Spitzer} and radio data over relatively small $<10$~square degree areas \citep[e.g.][]{Appleton04, Frayer06, Ibar08, Murphy09,Michalowski10, Sargent10,Bourne10}. This has led to several authors suggesting that the FIRC remains unchanged out to high redshift ($z\gtsim 1.5$) \citep[e.g.][]{Sargent10}, whereas others suggest a shallow decrease in the ratio between far-infrared luminosity and radio luminosity \citep[e.g.][]{Seymour09}. Constraining the evolution of the FIRC is important as it may help our understanding of the physical mechanism which results in such a tight correlation between the thermal and non-thermal emission. For example, metallicity and temperature evolution would effect the far-infrared luminosity but not necessarily the radio synchrotron emission, where as an evolution in the magnetic field strength would alter the radio emission. Furthermore, evolution of the gas density and the scale height of a galactic disk may also influence both emission mechanisms \citep[e.g.][]{HelouBicay93,NiklasBeck97}.

The Balloon-borne Large Aperture Submillimetre Telescope \citep[BLAST; ][]{Devlin09} experiment followed by the launch of the {\em Herschel Space Observatory} \citep{Pilbratt10} open up the possibility of fully investigating the FIRC using spectral energy distributions which have been measured with far-infrared and sub-millimetre photometry. The first of these concentrated on the relatively deep fields of the extended {\em Chandra} Deep Field South with BLAST \citep{Ivison10a} and GOODS-North with {\em Herschel}  \citep{Ivison10b}. Both of these studies find evidence for a modest evolution in the FIRC from low-redshift through to the high-redshift Universe, although if confined solely to the far-infrared selected galaxies in these samples the evidence for evolution is not as significant.

However, we are still lacking a statistically significant sample of objects which can be used to determine the form of the FIRC in the low-redshift ($0.05<z<0.2$) Universe. \citet{Yun01} used data from the {\em Infrared Astronomical Satellite} \citep[{\em IRAS}; ][]{Neugebauer84} combined with the NRAO VLA Northern Sky Survey \citep[NVSS;][]{Condon98} to investigate the FIRC in the very low-redshift Universe (i.e. $z<0.05$) and found that in their sample of 1809 galaxies $\geq 98$~per cent were consistent with a simple linear FIRC. The next step is obviously to push beyond these very low redshifts and to fill the gap in the measurements of the FIRC between the \citet{Yun01} sample at $z<0.05$ and those conducted on smaller fields at higher redshifts ($z> 0.5$). 

We are now in a position to do this with the data acquired as part of the {\em Herschel} Astrophysical Tera-Hertz Large Area Survey \citep[H-ATLAS;][]{Eales10}. H-ATLAS is the largest open time key project on the {\em Herschel Space Observatory}. The survey aims to map 550~deg$^{2}$ at five wavelengths from $100-500~\mu$m to 5$\sigma$ point source sensitivities, including confusion noise, of $132, 126, 32, 36$ and $45$~mJy at 100, 160, 250, 350 and 500~$\mu$m respectively. 
The main scientific goals are to gain a complete understanding of galaxies in the relatively low-redshift Universe \citep[e.g.][]{Dye10, Amblard10}, to study the evolution of rare objects, such as active galactic nuclei \citep[AGN; ][]{Serjeant10, Hardcastle10,Bonfield10} and galaxy clusters, and also to act as a unique data set for finding distant lensed galaxies \citep{Negrello10}. 
In this paper we measure the FIRC using the Science Demonstration Phase of the H-ATLAS survey which covers $\sim 14$~degree$^{2}$ over an equatorial field centred at $09^h05^m30^s$, $+00^\circ30^{\prime}00^{\prime\prime}$, corresponding to one of the fields within the Galaxy and Mass Assembly \citep[GAMA; ][]{Driver09} Survey which has a wealth of other multi-wavelength data.  

The paper is arranged as follows: In Section~\ref{sec:sample} we discuss how our sample was selected using far-infrared data from {\em Herschel}, along with optical, near-infrared and radio data. In Section~\ref{sec:firc} we measure the FIRC and investigate how this may evolve with redshift and whether there is any dependence on either radio or far-infrared luminosity. In Section~\ref{sec:conc} we present our conclusions. Throughout the paper we use a concordance cosmology with $H_{0}=70$~km~s$^{-1}$~Mpc$^{-1}$, $\Omega_{\rm M} =0.3$ and $\Omega_{\Lambda} = 0.7$.

\section{Sample Selection}\label{sec:sample}

\begin{figure}
\includegraphics[width=1.\columnwidth]{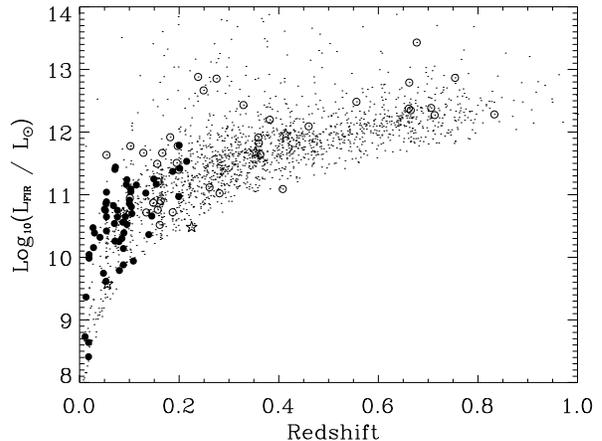}
\caption{Integrated far-infrared luminosity versus redshift for the $>5\sigma$ sources from the H-ATLAS SDP field. The small dots denote all of the H-ATLAS sources which we use for the stacking of the radio data (see Section~\ref{sec:stacking}), while the filled circles denote those sources with $>5 \sigma$ radio detections for which we can calculate the FIRC directly. The open circles represent those sources with $L_{1.4\rm GHz}> 10^{23}$~W~Hz$^{-1}$, i.e. in the AGN-luminosity regime, and the open stars are those objects with morphological signs of AGN activity in the radio maps (see Section~\ref{sec:firc}).}
\label{fig:L_z}
\end{figure}

For this study we use the H-ATLAS Science Demonstration Phase (SDP) data. Full details of the Spectral and Photometric Imaging Receiver \citep[SPIRE; ][]{Griffin10} and Photodetector Array Camera \citep[PACS; ][]{Poglitsch10} data reduction for the H-ATLAS data can be found in \citet{Pascale10} and \citet{Ibar10} respectively. The H-ATLAS 5$\sigma$ source catalogue \citep{Rigby10} was subsequently cross-matched with the Sloan Digital Sky Survey
Data Release 7 \citep[SDSS;][]{sdssdr7} and the United Kingdom Infrared Deep Sky
Survey Large Area Survey \citep[UKIDSS-LAS;][]{ukidss-lawrence}. Full details of the
cross-matching procedure can be found in \citet{Smith10}; however
to summarize, we use the SDSS DR7 $r$-band catalogue, limited to $r \le
22.4$, and perform a likelihood ratio analysis \citep{SutherlandSaunders92} on all possible counterparts within 10~arcseconds of the 250~$\mu$m
positions (the beamsize at 250~$\mu$m is 18~arcseconds). The likelihood ratio method relies on using not only
positional offsets, but also magnitude information -- both of the
counterparts and of the full catalogue -- to determine the ratio of
the probability that an optical source is related to the 250~$\mu$m
source to the probability that the optical source is unrelated. One
can then use these values to determine a reliability, $R$, which
estimates the probability that a given source is the true counterpart
to the {\em Herschel} object. The likelihood ratio accounts for the
fact that not all {\em Herschel} sources will be detected in the SDSS $r$-band data, due to many galaxies residing at higher redshift than the SDSS is sensitive to, and also possibly due to dust obscuration. In what follows, we only consider those 2334 sources with
reliable counterparts ($R > 0.8$) to the {\em Herschel}-selected catalogue. 

\subsection{Spectroscopic and Photometric redshifts}

For our study on the evolution of the FIRC, redshift information for all of the sources is crucial. For the brighter objects in our sample we are able to use the SDSS spectroscopic redshifts, for the fainter sources we use the year 1 and 2 data from the Galaxy and Mass Assembly (GAMA) survey \citep{Driver10} where available. The target selection for the GAMA survey was $r < 19.4$ or $z < 18.2$ or $K_{\rm AB} < 17.6$, full details of the target selection and the target priorities can be found in \citet{Baldry10} and \citet{Robotham10} respectively.  The combination of the SDSS and GAMA redshift surveys results in 877 sources having spectroscopic redshifts.

Using the combination of SDSS $ugriz$ and UKIDSS-LAS $YJHK$ data we determine photometric redshifts for the sources without spectroscopic redshifts. We used the publicly available {\sc{annz}} \citep{annz04}  photometric redshift code, using spectroscopic redshifts from
the SDSS, GAMA \citep{Driver10}, DEEP2 \citep{deep2-2007}, zCOSMOS \citep{zcosmos} and 2SLAQ \citep{2slaq-lrg}. This allows us to
construct a spectroscopic training set with large numbers of objects
($>1000$ per bin of unit magnitude or 0.1 in redshift) up to $r$-band
magnitudes $r<23$ and redshifts $z<1.0$, i.e. to approximately the
photometric depth of SDSS and UKIDSS-LAS. 
Further details of the photometric redshifts can be found in \citet{Smith10}. For our analysis we always use the spectroscopic redshift, where available, in preference to the photometric redshifts which have a typical error of $\Delta z/(1+z) \sim 0.03$. In all cases where we has a spectroscopic redshift, this agreed with the photometric redshift within the uncertainties.


\subsection{The radio data}
Based on this {\em Herschel} catalogue, with optical identifications, we measured radio fluxes directly from the imaging data of both the NVSS and the Faint Images of the Radio Sky at Twenty-one centimetres \citep[FIRST;][]{Becker95} at the position of the optical galaxy. The NVSS has a typical rms flux-density of 0.45~mJy and has the benefit that it was carried out in VLA D-Array and therefore has a resolution of around $\sim 45$~arcsec which is larger than the expected size of any individual galaxy within the H-ATLAS galaxy sample. As such we expect little flux to be missed or resolved out using the NVSS. On the other hand the FIRST survey with a 5-arcsec beam is deeper than NVSS (rms = 0.15~mJy) but may
miss extended flux, particularly at the low-redshift end of our source distribution. We therefore use the NVSS survey to measure the radio fluxes at the positions of the H-ATLAS sources with spectroscopic or photometric redshifts $z< 0.2$ and the FIRST survey to measure the radio fluxes of sources with $z>0.2$, where the FIRST beam equates to a galaxy size of $\gtsim 16$~kpc.  We include all sources with a flux-density $S_{1.4 \rm GHz} > 5$ times the background noise level determined using each individual $5\times 5$~arcmin$^{2}$ cut-out around each source. We note that only two of the sources with direct radio detections at $> 5\sigma$ lie above a redshift of $z=0.2$ Therefore this split makes little difference to the results presented in this paper, but may be important when comparing to results which investigate the FIRC at higher redshifts. To summarise, our sample is comprised of all objects detected at $\geq 5\sigma$ in the H-ATLAS catalogue, with radio counterparts detected at $> 5\sigma$ above the local background rms from the NVSS ($z< 0.2$) and FIRST ($z>0.2$) radio surveys. It is possible that the lower resolution NVSS survey could suffer from some source confusion, where multiple sources contribute to the measured flux of the object. We checked for this using the higher resolution FIRST data to identify whether there was more than one source within the NVSS beam, and where this was deemed to be an issue ($< 5$~per cent) the FIRST maps was used to measure the flux. We also checked for extended emission associated with AGN as detailed in Section~\ref{sec:firc}.

\subsection{Far-infrared luminosities}
Far-infrared luminosities were derived using two methods. The first uses a simple modified blackbody fit to the three SPIRE bands at 250, 350 and 500$\mu$m, and the PACS data at 100 and 160$\mu$m where available. We fixed the dust emissivity index to $\beta = 1.5$ and varied the temperature over the range $10<T<50$~K, following \citet{Dye10}. As in \citet{Dye10} we find a median temperature of $T=26$~K. For each source we measured the integrated far-infrared luminosity ($8-1000~\mu$m) and the rest-frame 250~$\mu$m luminosity. We also fit the data with an emissivity index of $\beta=2$, which gave a median temperature of $T=23$~K, again consistent with \citet{Dye10}. However, the integrated far-infrared luminosities derived for both fits were consistent within the uncertainties with a median offset of $+0.07$~dex in integrated far-infrared luminosity for $\beta=1.5$ compared to those fit with $\beta =2$. 

It is widely acknowledged that simple modified blackbody fits may underestimate the total far-infrared emission, particularly at the short wavelength end of the spectrum. For this reason we also determine the far-infrared luminosity of our sources using the comprehensive set of star-burst  SEDs from \citet{SK07}, in a similar way to \citet{Symeonidis08}. All star-burst and Ultra-luminous Infrared Galaxy (ULIRG) SEDs were considered to fit the data\footnote{see http://www.eso.org/$\sim$rsiebenm/sb\_models}. We use all three SPIRE data points and the PACS detections where available. In the absence of a detection at the PACS wavebands we use the $2\sigma$ upper limit to constrain the SED to lie below these values. We note that no shorter wavelength data is available for the H-ATLAS field at the present time.
We find that the templates provide far-infrared luminosities approximately 0.7~dex more luminous than the corresponding modified blackbody fits, although the monochromatic luminosities at the rest-frame wavelengths covered by the SPIRE data are consistent, as expected. For the purposes of this paper we use the far-infrared luminosities determined from the full SED fits using the templates from \citet{SK07} and for the monochromatic luminosities we use the modified blackbody fits, although we emphasise that these are consistent between the two methods.

 Fig.~\ref{fig:L_z} shows the distribution of our sources on the far-infrared luminosity versus redshift plane, which demonstrates that we are able to sample the $z<0.2$ Universe sufficiently well for individual sources to investigate the form of the FIRC with this sample. We use stacking of the radio data (Sec.~\ref{sec:stacking}) to determine whether our findings at $z<0.2$ are consistent with the average sources properties up to $z\sim 0.5$ and beyond, by comparing with previous results in the literature \citep[e.g.][]{Ivison10a,Ivison10b,Bourne10}.


We also use the results described in \citet{Hardcastle10} to obtain the far-infrared luminosities for those sources which are strongly detected at radio wavelengths but which fall below the $5\sigma$ detection limit of our {\em Herschel} catalogue, and thus do not make it into our sample. The combination of these allows us to probe to lower and higher radio luminosities at all redshifts, thus negating some of the biases inherent to selecting in a single band. We show these for completeness; however, we do not use them in calculating the FIRC.

\begin{figure}
\includegraphics[width=1.\columnwidth]{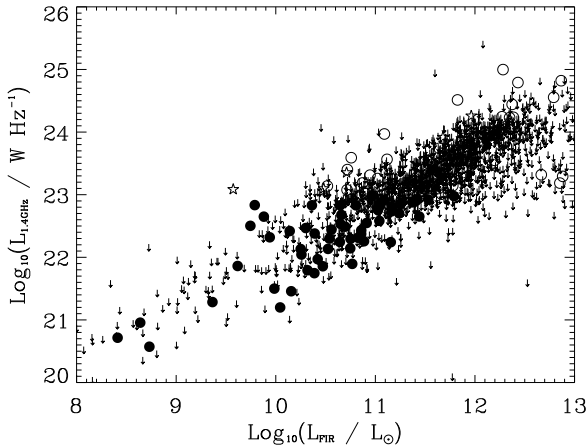}
\caption{Integrated far-infrared luminosity versus rest-frame 1.4~GHz radio luminosity of the high-reliability 250~$\mu$m detected sources in the H-ATLAS SD field with radio cross-identifications $> 5\sigma$ above the local noise in the NVSS and FIRST images. NVSS fluxes were used for the sources at redshifts $z < 0.2$ and FIRST for those sources with $z > 0.2$, as explained in the text. The open stars represent those sources with morphological signs of AGN activity (see Section~\ref{sec:firc}) and the open circles denote those sources with $L_{1.4\rm GHz} > 10^{23}$~W~Hz$^{-1}$, and thus have a high probability of AGN contamination in their radio emission. The arrows denote the $5\sigma$ upper limits in the radio luminosity from the NVSS and FIRST. }
\label{fig:Lrad_vs_Lfir}
\end{figure}

In Fig.~\ref{fig:Lrad_vs_Lfir} we show the correlation between the far-infrared luminosity and the rest-frame 1.4~GHz radio luminosity. We assume a radio spectral index of $\alpha =0.8$\footnote{We use the definition of spectral index such that the radio flux density $S_{\nu} \propto \nu^{-\alpha}$, where $\nu$ is the frequency.}, as found for submillimetre galaxies by \citet{Ibar10a}, for the 104 objects (72 with spectroscopic redshifts and 32 with photometric redshifts) which lie above our flux limit in H-ATLAS and with $>5$~$\sigma$ detections in the radio maps. One can easily see the strong correlation between the far-infrared and radio emission in our sample, with the best fit relation $\log_{10}L_{\rm FIR} = (17.8 \pm 0.7) + (0.42 \pm 0.07)\log_{10}L_{\rm 1.4GHz}$ for those sources with $L_{\rm 1.4~GHz} < 10^{23}$~W~Hz$^{-1}$ (see Section~\ref{sec:firc}). 

\subsection{Stacking}\label{sec:stacking}
For those sources detected at {\em Herschel} wavelengths but which fall below our 5$\sigma$ threshold from the radio data, we also perform a stacking analysis on the NVSS and FIRST radio images to obtain statistical detections. To do this we followed the technique of \citet{White07}, which clips the image at 5$\sigma$ around the median value and then combines the individual images weighted by their variance \citep[see also][]{Falder10}. The stacked images are generated in various bins of far-infrared luminosity and redshift to allow direct comparisons to those sources with direct detections. Again we do not use these to derive the median and mean values for the FIRC but just as a guide so the reader can judge for consistency.

\section{The far-infrared radio correlation}\label{sec:firc}

The FIRC we use here is defined as the logarithmic ratio of the integrated far-infrared flux, $S_{\rm IR}$, determined from rest-frame wavelengths $8-1000$~$\mu$m \citep[following ][]{Bell03,Ivison10a}, and the rest-frame 1.4~GHz $k$-corrected flux density, $S_{1.4\rm GHz}$ such that  $q_{\rm IR} = \log_{10}[(S_{\rm IR}/3.75 \times 10^{12})/(S_{\rm 1.4GHz})]$ and is dimensionless. $S_{\rm IR}$ has units of W~m$^{-2}$, $3.75\times10^{12}$~Hz is the normalising frequency (Helou et al. 1985) and $S_{1.4\rm GHz}$ has units of W~m$^{-2}$~Hz$^{-1}$.  Following \citet{Ivison10a}, we also calculate the monochromatic relationship between the $k$-corrected emission at $250\mu$m (using our modified blackbody fit) and 1.4~GHz (using a spectral index of $\alpha=0.8$), hereafter $q_{250}$.

Contamination in the sample from radio-luminous AGN would preferentially decrease the value of $q_{\rm IR}$. The radio luminosity at which AGN begin to dominate the source population occurs around $L_{\rm 1.4GHz} \gtsim 10^{23}$~W~Hz$^{-1}$ \citep[e.g.][]{MauchSadler07, Wilman08}. 
Inspecting the FIRST radio images we find that nine of the sources in our sample show clear signs of AGN activity, i.e. jet-like phenomena. All of these sources have radio luminosities $L_{\rm 1.4GHz} > 10^{23}$~W~Hz$^{-1}$ and $q_{\rm IR}$ values below the sample mean, as would be expected for an AGN dominated source. We omit these nine sources from all subsequent analyses, even though the effect they would have on our results is negligible. All objects with radio luminosities $L_{\rm 1.4~GHz} > 10^{23}$~W~Hz$^{-1}$ are represented with open symbols in all figures.

\subsection{Redshift dependence of the far-infrared--radio correlation}\label{sec:zdependence}

\begin{figure}
\includegraphics[width=1.\columnwidth]{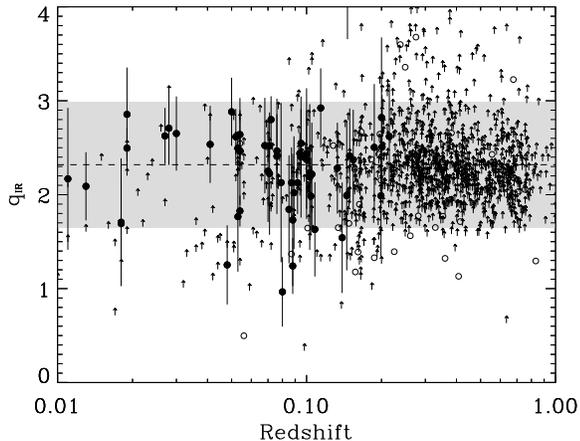}
\caption{The FIRC versus redshift for our complete sample. Solid circles denote the sources with $L_{\rm 1.4GHz} \leq 10^{23}$~W~Hz$^{-1}$, whereas the open circles represent the sources with $L_{\rm 1.4GHz} > 10^{23}$~W~Hz$^{-1}$, i.e. the radio luminosity at which AGN begin to dominate the source population. The dashed line is the weighted mean value of $q_{\rm IR}$ for the $L_{\rm 1.4GHz} < 10^{23}$~W~Hz$^{-1}$  sources and the grey shaded region shows the standard deviation on $q_{\rm IR}$. The arrows denote the $5\sigma$ lower limits based on the upper limits from the radio data. }
\label{fig:qzall}
\end{figure}

In Fig.~\ref{fig:qzall} we show the $q_{\rm IR}$ parameter as a function of redshift for all of the objects in our sample with $>5\sigma$ detections at 1.4~GHz. If we consider this entire sample then we obtain a mean value of $q_{\rm IR} = 2.15 \pm 0.09$ and median $q_{\rm IR} = 2.22 \pm 0.11$.
We also measure the mean and median values of $q_{\rm IR}$ after excluding the radio sources with obvious signs of AGN activity {\em and} those above $L_{\rm 1.4~GHz} > 10^{23}$~W~Hz$^{-1}$. Unsurprisingly this leads to a larger value of the mean, $q_{\rm IR} = 2.33 \pm 0.09$ and median $q_{\rm IR} = 2.40 \pm 0.12$. To estimate the effect of the lower limits in our measurement of the value for $q_{\rm IR}$ we use the Kaplan-Meier estimator \citep{FeigelsonNelson85}, which is a maximum likelihood estimator to calculate the mean in the presence of lower or upp limits \citep[see e.g.][]{Sajina08}. Using this method we find a mean of $q_{\rm IR} = 2.52 \pm 0.03$.
These are consistent with the value of $q_{\rm IR}$ measured in the very local Universe \citep{Yun01} using {\em IRAS} data, who found $q_{\rm IR}=2.34\pm 0.01$, although we note that \citet{Yun01} calculated the value of $q_{\rm IR}$ based on the $40<\lambda<120$~$\mu$m wavelength range only. \citet{Bourne10} also discuss this point and suggest that the difference between the value of $q_{\rm IR}$ measured in this limited wavelength ranged compared to the total infrared luminosity is $\sim 0.3$~dex for an M51 template SED, but note that this is subject to large variations due to different relative contribution from cooler dust. We use our models to find what range this correction spans and we also find a median difference of 0.3~dex with a standard deviation of 0.4~dex in the fra-infrared luminosity; thus the consistency remains at the 2~$\sigma$ level with the results of \citet{Yun01}. Comparing with the results of \citet{Bell03}, who used the bolometric far-infrared luminosity of local galaxies, and found $q_{\rm IR} = 2.64$, we find that our results are slightly lower, but again within 2~$\sigma$.
Our value of $q_{\rm IR}$ is also in good agreement with the values measured at higher redshift by various authors \citep[e.g.][]{Ivison10a,Ivison10b,Sargent10} who find values in the range $2.4 < q_{\rm IR} < 2.8$. Some find evidence for an increase in $q_{\rm IR}$ towards low redshift \citep[e.g.][]{Bourne10}, but our data do not support this; we return to this point in Section~\ref{sec:binrad}.

\begin{figure}
\includegraphics[width=1.\columnwidth]{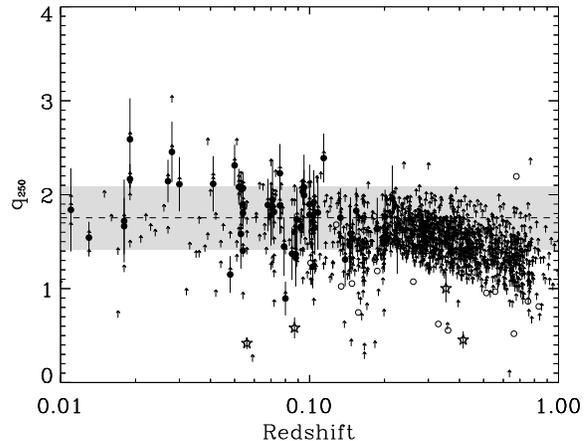}%
\caption{The $q_{250}$--radio correlation as a function of redshift. The open circles denote those sources with $L_{\rm 1.4~GHz} > 10^{23}~$W~Hz$^{-1}$. The dashed line is the weighted mean value of $q_{\rm 250}$ for the $L_{\rm 1.4GHz} < 10^{23}$~W~Hz$^{-1}$  sources, and the grey shaded region shows the standard deviation on $q_{\rm 250}$. The arrows denote the $5\sigma$ lower limits based on the upper limits from the radio data.}
\label{fig:250z}
\end{figure}

Fig.~\ref{fig:250z} shows the value of $q_{250}$ against redshift for our sample. Only considering those objects with $L_{\rm 1.4GHz} < 10^{23}$~W~Hz$^{-1}$, we determine a mean $q_{250}=1.78 \pm 0.04$ and median $q_{250} = 1.84\pm 0.05$, similar to the values quoted by \citet{Ivison10a} with the $k$-correction applied. Using the lower limits and the Kaplan-Meier estimator we find a mean $q_{250} = 2.01\pm 0.04$.
However, we find that the dispersion on $q_{250}$ is a factor of two tighter than the corresponding dispersion for $q_{\rm IR}$. This suggests that the flux measured at rest-frame 250~$\mu$m is possibly more closely related to the radio emission than the integrated far-infrared luminosity, although we note that the larger dispersion could be due to the large uncertainties of the complete SED due to our lack of direct detections in the PACS observations, particularly for the sources lying close to the flux-density limit at SPIRE wavelengths.

However, there are reasons why we might expect to see a larger dispersion in $q_{\rm IR}$ than in $q_{250}$. For example, \citet{Dunne00b} showed that the relationship between the radio emission and the 850$\mu$m emission measured with SCUBA has significantly more scatter than in the similar relationship between radio emission and the far-infrared emission determined from a fit to the SED which includes data from {\em IRAS}. Similarly, \citet{Vlahakis07} showed that the relation between  850$\mu$m emission and radio emission has significantly more scatter than the relationship between 60$\mu$m and 1.4~GHz. Both suggest that this could be due to the presence of cold dust which dominates the emission at sub-millimetre wavelengths, but which is not as strongly associated with massive star formation. Given that our total far-infrared luminosities are largely constrained by the longer wavelength SPIRE data points, then we would expect the scatter in these value to be larger than the scatter in monochromatic luminosities derived at 250~$\mu$m which in this scenario would more closely trace the dust heated by young stars.

This explanation can also be reconciled with detailed studies of individual galaxies. Using {\em Herschel} observations of M81, \citet{Bendo10} recently found evidence that the longer wavelength (160-500$\mu$m) emission traces dust with temperature 15-30~K which is heated by evolved stars, whereas the shorter wavelength ($\sim 70\mu$m) emission is dominated by dust heated by young stars in star-forming regions.
Furthermore, in a sample of nearby galaxies observed with {\em Spitzer}, \citet{Calzetti10} also found evidence that as one moves to longer wavelengths, the conversion to a star-formation rate becomes more uncertain with the dispersion increasing by a factor of $\sim2$ moving from 24$\mu$m to 70$\mu$m, although \citet{Appleton04} find the opposite to be true, this could possibly due to the wide redshift range, $0<z<1$ (and thus luminosity range) in the Appleton et al. study. Thus, it appears that we may be seeing this effect on the FIRC, where the radio emission remains a robust tracer of the star-formation rate but the longer wavelength {\em Herschel} data has an increasing contribution to the emission from dust heated by evolved stellar populations, compared to the heating by young massive stars \citep[see also][]{MRR03, Bell03}.

\subsection{Luminosity dependence of the far-infrared--radio correlation}
\begin{figure*}
\includegraphics[width=1.5\columnwidth]{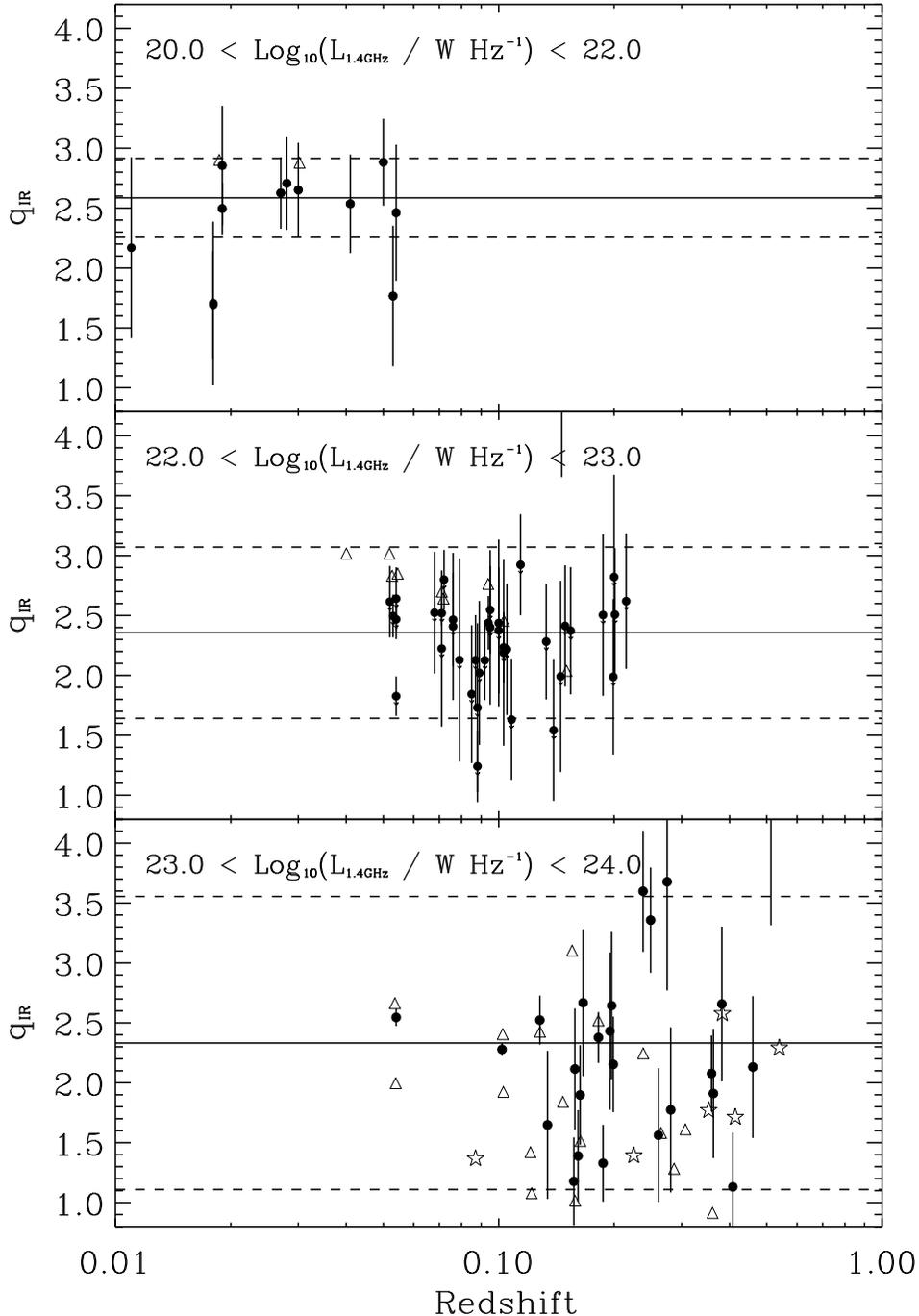}%
\caption{The FIRC as a function of redshift separated into three radio luminosity bins shown in the individual panels. The solid circles denote the 5$\sigma$ detections at the radio wavelength, the open stars denote those sources identified as AGN on the basis of jet-like radio structure and the open triangles denote those object selected at radio wavelengths from \citet{Hardcastle10}. The solid horizontal line shows the weighted mean value of $q_{\rm IR}$ for all non-AGN sources with $L_{\rm 1.4GHz} < 10^{23}$~W~Hz$^{-1}$, and the dashed horizontal lines are the $1\sigma$ standard deviation of the same sources.}
\label{fig:qz1}
\end{figure*}

\begin{figure*}
\includegraphics[width=1.5\columnwidth]{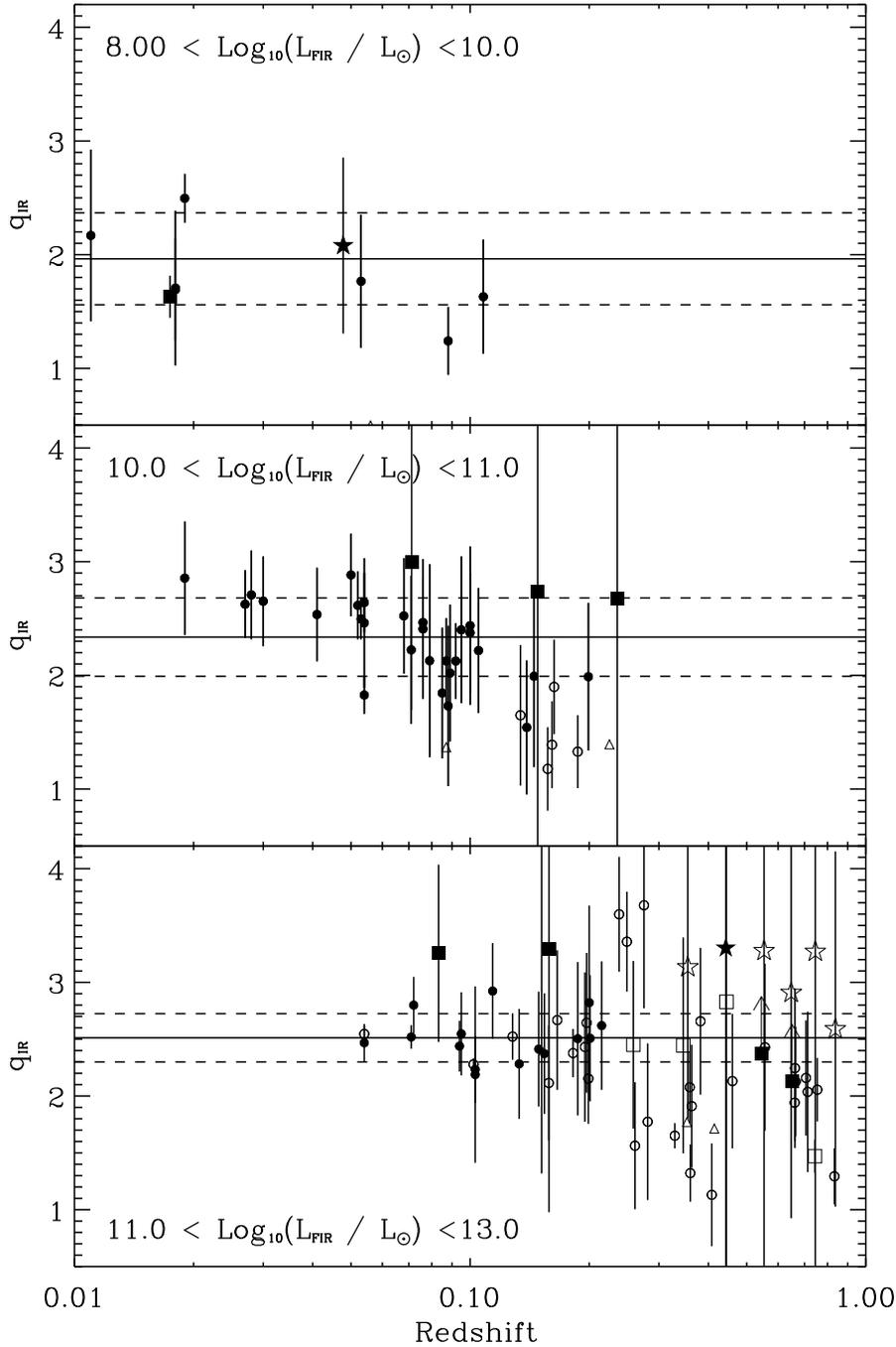}%
\caption{The FIRC as a function of redshift separated into three far-infrared luminosity bins shown in the individual panels. The solid circles denote the 5$\sigma$ detections at the radio wavelength with $L_{\rm 1.4GHz} < 10^{23}$~W~Hz$^{-1}$  and the open circles denote $5\sigma$ radio sources with $L_{\rm 1.4GHz} > 10^{23}$~W~Hz$^{-1}$. The open stars denote those sources identified as AGN on the basis of jet-like radio structure. 
We also show the results of our stacking analysis of the radio data at the position of the {\em Herschel} sources. The large filled squares and stars denote the average $q_{\rm IR}$ value from stacking at particular redshifts and far-infrared luminosities. In the top panel the filled square represents the mean $q_{\rm IR}$ value at $8< \log_{10}(L_{\rm FIR} / L_{\odot}) < 9$ and the filled star  $9< \log_{10}(L_{\rm FIR} / L_{\odot}) < 10$. In the middle panel the filled square denotes the value of $q_{\rm IR}$  from the stacked radio data for  $10< \log_{10}(L_{\rm FIR} / L_{\odot}) < 11$ plotted at the mean redshift of the sources. Similarly in the bottom panel the filled square represents the stacked data at $11< \log_{10}(L_{\rm FIR} / L_{\odot}) < 12$ and the filled star $12< \log_{10}(L_{\rm FIR} / L_{\odot}) < 13$ .
The open squares and stars represent the same luminosity range but for radio stacks which give flux-densities such that $L_{\rm 1.4GHz} > 10^{23}$~W~Hz$^{-1}$, i.e. within the AGN-dominated radio regime. As in Fig.~\ref{fig:qz1} the solid line shows the weighted mean value of $q_{\rm IR}$ for each far-infrared luminosity bin, with the dashed lines showing the standard deviation of the data for $L_{\rm 1.4GHz} < 10^{23}~$W~Hz$^{-1}$.}
\label{fig:qz2}
\end{figure*}

With the large number of objects available to us due to the high-sensitivity of the {\em  Herschel Space Observatory}, we are able to investigate the form of the far-infrared radio correlation over a range of redshifts, far-infrared luminosities and radio luminosities. However,  both the far-infrared data and the radio data are derived from flux-density limited parent images and any analysis is susceptible to Malmquist bias, where the higher redshift sources are biased towards the most luminous sources, and Eddington-type bias for signal-to-noise limited catalogues. Therefore, following both \citet{Sargent10} and \citet{Ivison10b} we split our sample into three radio luminosity bins (Fig.~\ref{fig:qz1}) and three far-infrared luminosity bins (Fig.~\ref{fig:qz2}) and determine the mean and median value for $q_{\rm IR}$ in each of these bins. 

\begin{figure*}
\includegraphics[width=1.5\columnwidth]{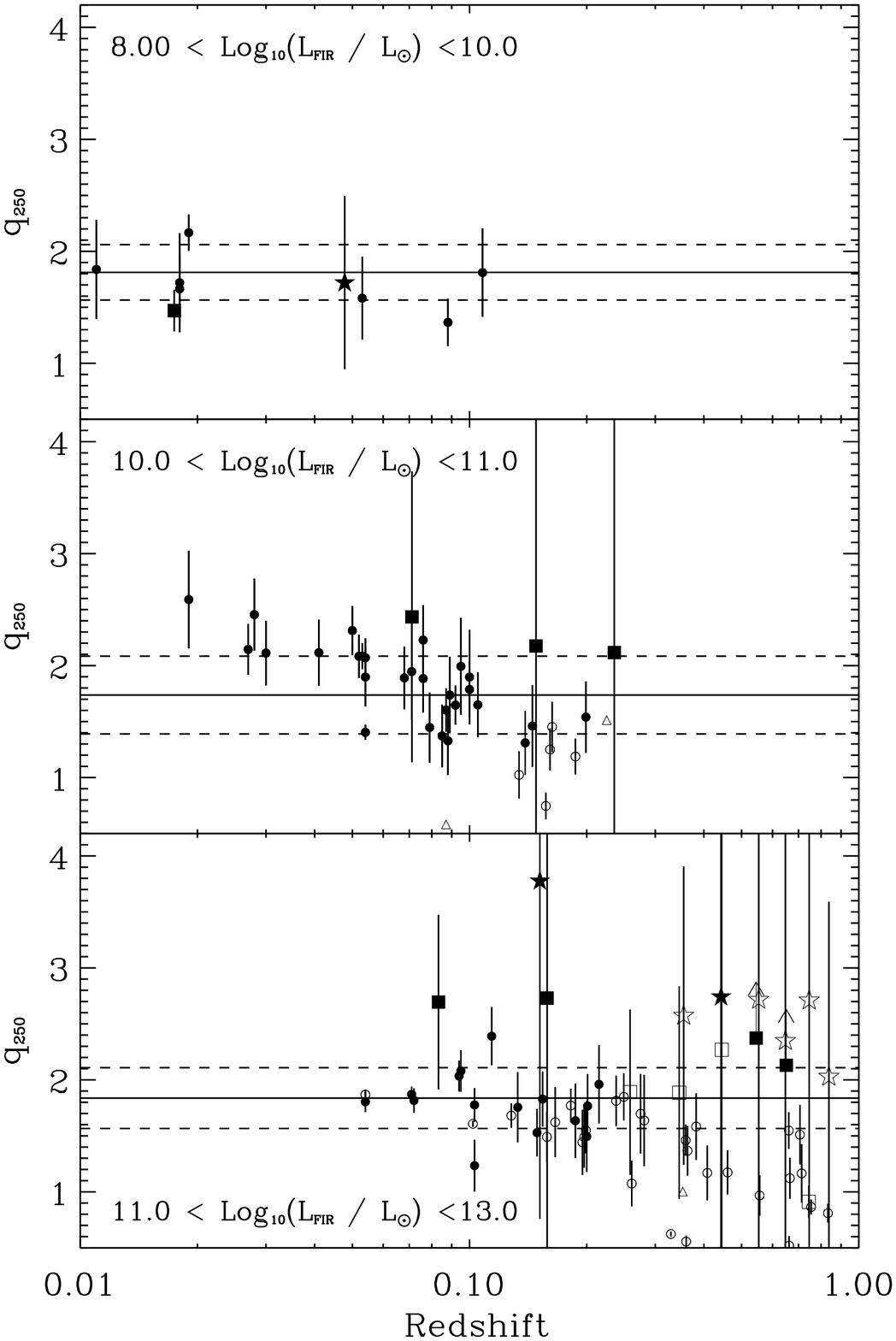}%
\caption{$q_{250}$ as a function of redshift separated into three far-infrared luminosity bins shown in the individual panels. The symbols are the same as in Fig.~\ref{fig:qz2}.}
\label{fig:q250split}
\end{figure*}

\subsubsection{Binning in radio luminosity}\label{sec:binrad}
Fig.~\ref{fig:qz1} shows that there is very little evidence for strong evolution in the FIRC over the redshift range $0<z<0.5$ in any of the individual radio luminosity bins, where a constant value of $q_{\rm IR}$ is consistent with the data. We also find similar results when using $q_{250}$.
However, we find that the mean (and median) values of $q_{\rm IR}$ ($q_{250}$) decrease with increasing radio luminosity from $q_{\rm IR} = 2.55 \pm 0.11$ ($q_{250} = 2.23 \pm 0.10$) for $L_{1.4\rm GHz} < 10^{22}$W~Hz$^{-1}$ to $q_{\rm IR} = 2.35 \pm 0.11$ ($q_{250} = 1.93 \pm 0.04$) for $10^{22} <$ ($L_{1.4\rm GHz}$ / W~Hz$^{-1}$)$ < 10^{23}$, through to $q_{\rm IR} = 2.33 \pm 0.24$ ($q_{250} = 1.81 \pm 0.06$) for $L_{1.4\rm GHz} > 10^{23}$W~Hz$^{-1}$. The most straightforward explanation of this is that there is an increasing fraction of  AGN as we push to higher radio luminosities. 

\begin{figure}
\includegraphics[width=1.\columnwidth]{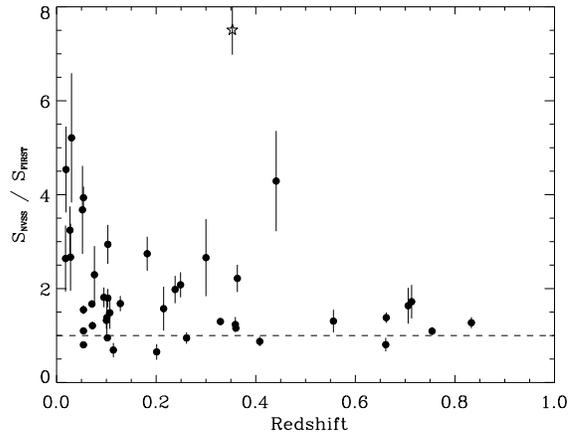}%
\caption{The ratio of the flux-densities from the NVSS and FIRST survey as a function of redshift for radio sources in our sample with $>4\sigma$ detections in both surveys. Those sources identified as AGN from their radio morphology are denoted by open stars. }
\label{fig:nvssfirst}
\end{figure}

Another possibility is that the increase in $q_{\rm IR}$ and $q_{250}$ towards low redshift could be caused by the increasing probability of resolving out extended structure with relatively high-resolution radio data. We do not believe that this is an issue with our sample as we have restricted ourselves to using the low-resolution NVSS at $z<0.2$ and only use the FIRST survey at the higher redshifts $z>0.2$. However, this could offer an explanation of the increase in $q_{\rm IR}$ towards low redshift seen with surveys targeting the high-redshift Universe. At high redshift ($z>0.5$) short baseline observations are not as important, due to the fact that distant star-forming galaxies are relatively compact \citep[e.g.][]{Muxlow05} and are fully resolved with longer baselines, whereas short baselines are needed for studying lower redshift galaxies which have larger angular extent on the sky. This extended emission would be resolved without short baselines and radio flux would be missed, this in turn would increase the FIRC to higher values.
In Fig.~\ref{fig:nvssfirst} we show the ratio of the flux-densities from NVSS and FIRST as a function of redshift. One can immediately see that at $z<0.2$ the FIRST survey has missing flux when compared to the NVSS. Such a selection effect could explain the recent results which suggest an increase in $q_{\rm IR}$ towards low ($z<0.5$) redshift \citep[e.g.][]{Bourne10}. Therefore, a combination of AGN contamination, and the resolving out of extended radio emission, could be responsible for the observed increase in $q_{\rm IR}$ towards low redshift based on deep field data which uses higher resolution radio imaging.

However, if we return to the explanation in Section~\ref{sec:zdependence} in which the longer wavelength emission has an increasing contribution from evolved stars \citep{MRR03, Calzetti10}, then we would expect the low-redshift sources to exhibit less correlation with the radio emission due to the fact that our data necessarily probe longer rest-frame wavelengths. This would not have a large effect within our dataset which is limited to $z<0.2$ for direct detections, however when considering the evolution from $z\sim 1$ through to $z\sim 0$, this could explain the increase in $q_{\rm IR}$ towards the low redshifts, where there is an increased relative contribution to the total far-infrared luminosity from dust heated by evolved stars.

We therefore suggest that the evidence for an increase in the FIRC towards low redshifts found by previous authors \citep[e.g.][]{Bourne10} can be explained by one or a combination of resolving-out extended radio emission in high-resolution radio data coupled with an increased contribution from dust heated by evolved stars at progressively lower redshifts.

\subsubsection{Binning in far-infrared luminosity}

Figs.~\ref{fig:qz2} and \ref{fig:q250split} show the $q_{\rm IR}$ and $q_{250}$ parameters as a function of redshift, split into bins of far-infrared luminosity, respectively. We find no evidence that the mean $q_{\rm IR}$ or $q_{250}$ increase with far-infrared luminosity, with all values between bins consistent within 1~$\sigma$. However, it is also apparent that there is evidence for a negative trend in $q_{\rm IR}$ and $q_{250}$ as a function of redshift in at least the most populated bin in $L_{\rm FIR}$ at $10 < \log_{10}(L_{\rm FIR} / $~L~$_{\odot}) < 11$. In addition, we find that the stacked radio data provide values for $q_{\rm IR}$ ad $q_{250}$ which closely resemble the values, within the uncertainties, on the individually detected sources, thus we can also be confident that we are not suffering any significant bias against sources falling below our 5$\sigma$ radio detection limit.


One explanation for this negative trend with redshift is that any enhanced contribution to the radio emission from AGN activity in these sources would also decrease the value of $q_{\rm IR}$ and $q_{250}$, and possibly increase the scatter. This obviously plays some part in the observed trend as the open circles in Fig.~\ref{fig:qz2}, denoting the sources with $L_{\rm 1.4~GHz} > 10^{23}$~W~Hz$^{-1}$, all lie below the mean values of $q_{\rm IR}$ and $q_{250}$. 
For those sources with  $L_{\rm 1.4~GHz} < 10^{23}$~W~Hz$^{-1}$ only a very minor enhancement of the radio flux would be needed to cause the observed offset. Using the radio continuum simulation of \citet{Wilman08} we determine how many low-luminosity AGN would be expected over the redshift range $0 < z<0.2$, where our sources with low values of $q_{\rm IR}$ reside, and also with $10^{20} <L_{\rm 1.4~GHz} / $W~Hz$^{-1} < 10^{22}$. We find that we expect $\sim 1-2$ radio-quiet AGN or low-luminosity FR-I type \citep{FanaroffRiley74} sources per square degree above the 5$\sigma$ radio flux-density limit of NVSS, thus in the SDP field we would expect 15--30 sources where an AGN is contributing to the radio emission at $L_{1.4\rm GHz} < 10^{23}$~W~Hz$^{-1}$. Thus it is entirely plausible that low-luminosity AGN could lower the value of $q_{\rm IR}$ for a small fraction of the sources in our sample. Furthermore, low-level AGN activity has been found to be relatively common in late-type galaxies \citep[e.g.][]{Filho06, DesrochesHo09}.
Follow-up observations across several wavebands could be used to test this, e.g. high-resolution radio maps or deep X-ray imaging.

\section{Conclusions}\label{sec:conc}

We have used the H-ATLAS Science Demonstration Phase data to investigate the evolution of the FIRC over the redshift range  $0<z<0.5$. Using a combination of the far-infrared data from H-ATLAS and the radio data from the FIRST and NVSS we find no evidence for evolution in the FIRC, with a median $q_{\rm IR} = 2.40\pm 0.12$ over the redshift range $0<z<0.5$. This is consistent with previous work at $z\sim 0$ \citep[e.g.][]{Yun01}, as well as studies focused on  much higher redshifts \citep[e.g.][]{Ibar08,Sargent10,Ivison10a,Ivison10b,Bourne10}, in addition to empirical models based on radio and far-infrared luminosity functions (e.g. Wilman et al. 2010).

Splitting the sample into bins of radio luminosity shows that in our far-infrared selected sample the FIRC is consistent with being constant with radio luminosity for sources restricted to the regime where star-forming systems dominate the source counts, i.e. those with $L_{1.4 \rm GHz} < 10^{23}$~W~Hz$^{-1}$. 
We find that the dispersion in $q_{\rm IR}$ is a factor of two higher than the dispersion for $q_{250}$.
This could be explained if the longer wavelength emission, i.e. in the sub-millimetre regime, was not produced from massive star formation and thus does not trace the same physical mechanism as the radio or the mid- to far-infrared emission. Such an effect has been found for lower redshift galaxies where the SEDs have been studied in detail \citep[e.g.][]{Bendo10,Calzetti10}. This would imply that when calculating star-formation rates using {\em Herschel} photometry it may be more accurate to restrict this to just using the shorter, rest-frame $<250$~$\mu$m, data.
However, we cannot entirely rule out the possibility of low-level AGN activity contributing to the radio emission and this would lower the value of $q_{\rm IR}$. We also suggest that the increased dispersion in the value of $q_{\rm IR}$ could be linked to an increasing uncertainty in the models, which are less well constrained than the monochromatic flux at 250~$\mu$m, where we have data close to the rest-frame wavelength of interest. Deeper observations with PACS and/or observations at submm wavelengths could test which explanation is the more important.

We also conclude that the increase in the ratio of far-infrared to radio luminosity towards low redshift found in previous work may be partly due to the relatively high spatial resolution of the radio data used. Such data would begin to resolve out extended emission from the large angular size galaxies which become more dominant in the low-redshift Universe. Future work combining the H-ATLAS survey with lower frequency Giant Metrewave Radio Telescope (GMRT) data at 325~MHz and the Low Frequency Array \citep[LOFAR; ][]{Morganti10} survey data will also allow us to investigate the FIRC over a larger range in redshift, due to the increase in depth of the radio data, and as a function of radio spectral index.  This may provide important clues as to the link between the synchrotron emission traced by the radio data and the thermal emission traced by the H-ATLAS data.

\section{Acknowledgments}
MJJ acknowledges support from an RCUK fellowship and MJH thanks the Royal Society for support. JSD acknowledges the support of the Royal Society via a Wolfson Research Merit award, and also the support of the European Research Council via the award of an Advanced Grant.
The Herschel-ATLAS is a project with Herschel, which is an ESA space observatory with science instruments provided by European-led Principal Investigator consortia and with important participation from NASA. The H-ATLAS website is http://www.h-atlas.org/
U.S. participants in {\it
Herschel}-ATLAS
acknowledge support provided by NASA through a contract issued from JPL.

This work used data from the UKIDSS DR5 and the SDSS DR7. The UKIDSS project is defined in \cite{ukidss-lawrence} and uses the UKIRT Wide Field Camera \citep[WFCAM; ][]{Casali07}. Funding for the SDSS and SDSS-II has been provided by the Alfred P. Sloan Foundation, the Participating Institutions, The National Science Foundation, the U.S. Department of Energy, the National Aeronautics and Space Administration, the Japanese Monbukagakusho, the Max Planck Society and the Higher Education Funding Council for England.

GAMA is a joint European-Australasian project based around a spectroscopic campaign using the Anglo-Australian Telescope. The GAMA input catalogue is based on data taken from the Sloan Digital Sky Survey and the UKIRT Infrared Deep Sky Survey. Complementary imaging of the GAMA regions is being obtained by a number of independent survey programs including GALEX MIS, VST KIDS, VISTA VIKING, WISE, Herschel-ATLAS, GMRT and ASKAP providing UV to radio coverage. GAMA is funded by the STFC (UK), the ARC (Australia), the AAO, and the participating institutions. The GAMA website is: http://www.gama-survey.org/ .

\end{document}